\documentclass[sigconf,screen]{acmart}

\usepackage{graphicx}
\usepackage{xcolor}
\usepackage{multirow}
\usepackage[inline]{enumitem}
\usepackage{subcaption}
\usepackage{sistyle}
\usepackage[ruled]{algorithm2e}
\usepackage{booktabs}
\usepackage{caption}
\SIthousandsep{,}
\usepackage{makecell}
\usepackage[english]{babel}
\usepackage{algorithmic}
\usepackage{tikz}
\usepackage{wrapfig}
\usepackage{acronym}
\usepackage{CJKutf8}
\usepackage{hyperref}
\hypersetup{
  colorlinks   = true,    
  urlcolor     = blue,    
  linkcolor    = blue,    
  citecolor    = red    
}

\newcommand{\heading}[1]{\vspace*{.5mm}\noindent\textbf{#1.}}

\AtBeginDocument{%
  \providecommand\BibTeX{{%
    \normalfont B\kern-0.5em{\scshape i\kern-0.25em b}\kern-0.8em\TeX}}}

\definecolor{lightred}{rgb}{1, 0.8, 0.8}

\makeatletter
\g@addto@macro\normalsize{%
  \abovedisplayskip 3pt plus1pt 
  \belowdisplayskip 3pt plus1pt
  \abovedisplayshortskip  0pt plus1pt%
  \belowdisplayshortskip  0pt plus1pt
}
\makeatother

\acrodef{CV}{computer vision}
\acrodef{IR}{information retrieval}
\acrodef{LLM}{large language model}
\acrodef{MDP}{Markov decision process}
\acrodef{NLP}{natural language processing}
\acrodef{NRM}{neural ranking model}
\acrodef{RL}{reinforcement learning}

\author{Hongru Song}
\orcid{0009-0000-8443-9499}
\author{Yu-An Liu}
\orcid{0000-0002-9125-5097}
\affiliation{
 \institution{State Key Laboratory of AI Safety, Institute of Computing Technology, Chinese Academy of Sciences}
 \institution{University of Chinese Academy of Sciences}
 \city{Beijing}
 \country{China}
}
\email{{songhongru24s, liuyuan21b}@ict.ac.cn}

\author{Ruqing Zhang}
\orcid{0000-0003-4294-2541}
\authornote{Ruqing Zhang is the corresponding author.}
\author{Jiafeng Guo}
\orcid{0000-0002-9509-8674}
\affiliation{
 \institution{State Key Laboratory of AI Safety, Institute of Computing Technology, Chinese Academy of Sciences}
 \institution{University of Chinese Academy of Sciences}
 \city{Beijing}
 \country{China}
}
\email{{zhangruqing,guojiafeng}@ict.ac.cn}

\author{Maarten de Rijke}
\orcid{0000-0002-1086-0202}
\affiliation{
 \institution{University of Amsterdam}
 \city{Amsterdam}
 \country{The Netherlands}
}
\email{m.derijke@uva.nl}

\author{Sen Li}
\orcid{0000-0002-2124-953X}
\author{Wenjun Peng}
\orcid{0000-0003-2392-4946}
\affiliation{
 \institution{Taobao \& Tmall Group of Alibaba}
 \city{Hangzhou}
 \country{China}
}
\email{lisen.lisen@alibaba-inc.com}
\email{pengwj@mail.ustc.edu.cn}

\author{Fuyu Lv}
\orcid{0000-0001-5918-093X}
\affiliation{
 \institution{Taobao \& Tmall Group of Alibaba}
 \city{Hangzhou}
 \country{China}
}
\email{fuyu.lfy@alibaba-inc.com}

\author{Xueqi Cheng}
\orcid{0000-0002-5201-8195}
\affiliation{
 \institution{State Key Laboratory of AI Safety, Institute of Computing Technology, Chinese Academy of Sciences}
 \institution{University of Chinese Academy of Sciences}
 \city{Beijing}
 \country{China}
}
\email{cxq@ict.ac.cn}

\renewcommand{\shortauthors}{Hongru Song et al.}

\copyrightyear{2027}
\acmYear{2027}
\setcopyright{cc}
\setcctype{by}

\acmConference[WSDM '27]{Proceedings of the Twentieth ACM International Conference on Web Search and Data Mining}{February 15--19, 2027}{Hong Kong, Hong Kong}
\acmBooktitle{Proceedings of the Twentieth ACM International Conference on Web Search and Data Mining (WSDM '27), February 15--19, 2027, Hong Kong, Hong Kong}
\acmDOI{10.1145/3862322.3863468}
\acmISBN{979-8-4007-3117-4/2027/02}



\begin{document}

\title[PROSPER: LLMs as Sparse Retrievers for First-Stage Product Search]{PROSPER: LLMs as Sparse Retrievers\texorpdfstring{\\}{ }for First-Stage Product Search}

\begin{abstract}
Product search is a crucial component of large-scale e-commerce platforms, where first-stage retrieval should achieve high recall while ensuring efficient online deployment. 
Sparse retrieval remains attractive for this setting due to its interpretability and deployability, yet it suffers from severe vocabulary mismatch challenges, especially for short queries and product titles. 
Large language models (LLMs) offer strong semantic capabilities, but directly applying them to sparse retrieval in product search introduces new challenges: 
\begin{enumerate*}[label=(\roman*)] 
\item semantic expansion may distort critical literal signals such as brand names and model identifiers; and 
\item learning meaningful sparse representations is difficult in the high-dimensional vocabulary space induced by LLMs. 
\end{enumerate*} 

To address these challenges, we propose PROSPER, a framework that uses LLMs as sparse retrievers for product search while explicitly preserving lexical fidelity. 
PROSPER combines 
\begin{enumerate*}[label=(\roman*)] 
\item a literal residual network that alleviates lexical expansion hallucination by reinforcing underweighted literal terms through a residual compensation mechanism; and
\item a lexical focusing window that uses top-$k$ selection for rapid early sparsification and effective training initialization. 
\end{enumerate*}
Extensive offline evaluations and large-scale online A/B tests demonstrate that PROSPER consistently improves over existing sparse retrievers, and provides revenue gains for industrial hybrid retrieval systems. 
\end{abstract}

\maketitle

\section{Introduction}
E-commerce product search connects users with billion-scale product catalogs while processing hundreds of millions of daily queries under strict latency and memory constraints. 
Industrial retrieval pipelines typically follow the ``index-retrieve-then-rank'' \cite{10.1145/3637528.3671654} paradigm, where first-stage retrieval plays a key role by establishing the upper bound of achievable recall, i.e., relevant products missed at this stage cannot be recovered downstream. 

\heading{Dense retrieval: effectiveness at the cost of deployability} Dense retrieval has emerged as the dominant paradigm for first-stage retrieval, using neural encoders to map queries and documents into low-dimensional semantic spaces \cite{guoSemanticModelsFirststage2022,karpukhin-etal-2020-dense, 10.1145/3397271.3401075}. 
Despite their effectiveness, dense representations are difficult to interpret, hindering failure diagnosis in commercial systems. Dense indices also incur substantial storage and serving overheads and can be several times larger than the original corpus \cite{10.1145/3626772.3657951,xuCSPLADELearnedSparse2025}. These limitations motivate retrieval methods that better balance effectiveness, efficiency, and interpretability.

\heading{Sparse retrieval: efficiency and interpretability, but limited semantics} Sparse retrieval represents queries and documents as high-dimensional but sparse vectors aligned with lexical vocabularies, naturally supporting efficient inverted indexing and transparent matching \cite{9357643,10.1561/1500000019,10.1145/3404835.3463098}. 
They therefore remain a core component of industrial search systems. Learned sparse retrievers, most notably SPLADE \cite{10.1145/3269206.3271800,10.1145/3397271.3401204,10.1145/3404835.3463098}, jointly learn lexical expansion and term weighting, improving over classical statistical approaches. However, their masked language model backbones still struggle with vocabulary mismatch in product search, where short queries and titles are dominated by literal terms such as brands and model identifiers, leaving a persistent gap compared to dense retrievers.

\heading{LLMs as sparse retrievers: opportunities and challenges} Large language models (LLMs) offer stronger semantic representations and broader world knowledge than prior encoder architectures \cite{qwen2025qwen25technicalreport,taoLLMsAreAlso2024,minaee2025largelanguagemodelssurvey}, making them promising backbones for combining semantic understanding with efficient and interpretable sparse representations.
However, our preliminary analysis (Section~\ref{subsec:Application and Challenges}) reveals that naively replacing BERT with an LLM in existing sparse frameworks is ineffective and even detrimental in product search. 
Two fundamental challenges arise: 
\begin{enumerate*}[label=(\roman*)]
\item Semantic expansion driven by LLMs often distorts critical lexical signals: expanded terms may dominate the sparse representation, while exact-match terms, such as brands or model numbers that are essential for user intent, are underweighted.
\item The vastly enlarged vocabulary of LLMs makes training unstable and inefficient. Learning sparse representations for short queries and titles in a very high-dimensional space without guidance leads to poor initialization and slow convergence. 
\end{enumerate*}

\heading{PROSPER: Literal-preserving LLM-based sparse retrieval} 
To address these challenges, we propose PROSPER, a framework for PROduct search using LLMs as SParsE Retrievers while explicitly preserving lexical fidelity. 
PROSPER contributes at both the architectural and training levels: 
\begin{enumerate*}[label=(\roman*)]
\item We design a \emph{literal residual network} (LRN) that employs a compensatory weighting mechanism to amplify the importance of literal terms in user queries and items. This effectively mitigates irrelevant term generation (which we term \emph{lexical expansion hallucination}) by anchoring the model's attention to user-critical tokens, such as brand and model names. 
\item We apply top-$k$ sparsification~\cite{yang2021topk,xie2026adasparse} through a \emph{lexical focusing window} (LFW) and find it effective for product search, improving retrieval and accelerating early sparsification. As representations become sparse, the window becomes inactive and FLOPS regularization continues to refine term weights.
\end{enumerate*}
Overall, we present an empirical study of the challenges and practical solutions for LLM-based sparse retrieval in product search.

\heading{Experimental findings}
We evaluate PROSPER on both the Multi-CPR E-commerce dataset and a real-world dataset sampled from search engine logs.
On Multi-CPR E-commerce, PROSPER improves Hit@1000 by 10.2, 4.3, and 1.5 percentage points over BM25\cite{10.1561/1500000019}, SPLADE-v2 \cite{formalSPLADEV2Sparse2021}, and an LLM-based sparse retrieval baseline \cite{xuCSPLADELearnedSparse2025}, respectively.
Moreover, large-scale online A/B tests demonstrate consistent gains in key business metrics. In direct-guided settings, PROSPER improves net GMV by 1.60\%, net transactions by 0.31\%, and net UV by 0.24\%, further validating the practical value of PROSPER in real-world industrial search systems.
\vspace{-2mm}

\section{Preliminaries}
\label{sec:preliminary}
We recall SPLADE, the foundation of our method, and then discuss the challenges of applying LLM-based sparse retrieval to product search.

\vspace{-3mm}
\subsection{SPLADE}
\heading{Model architecture} Given an input query or document sequence (after WordPiece tokenization) $S = (t_1, t_2, \ldots, t_N)$ and its BERT embeddings $\mathbf{h} = (h_1, h_2, \ldots, h_N)$, SPLADE uses the masked language modeling head to project each hidden state to a vocabulary-sized vector $\mathbf{H}_i \in \mathbb{R}^{|V|}$. Its $j$-th dimension is the logit measuring the importance of vocabulary token $j$ to input token $i$. ReLU activation, log-saturation, and max-pooling across token positions produce the final representation:
\begin{equation}
    w_j = \max_{1 \le i \le N} \log(1 + \text{ReLU}(w_{ij})),
\end{equation}
where $w_{ij}$ represents the importance of vocabulary token $j$ for input token $i$, and $w_j$ is the final weight for term $j$ in the representation.

\heading{Model training} SPLADE combines ranking and regularization losses. Its ranking loss uses in-batch negatives~\cite{lin-etal-2021-batch,qu-etal-2021-rocketqa} and hard negatives \cite{xiong2020approximatenearestneighbornegative}. Given a query $q_i$, positive document $d_i^+$, hard negative $d_i^-$ (e.g., sampled by BM25), and in-batch negatives $\{d_{i,j}^-\}_j$, the InfoNCE loss \cite{oord2019representationlearningcontrastivepredictive} is:
\begin{equation}
    \mathcal{L}_\mathit{rank\text{-}IBN} = -\log \frac{e^{s(q_i, d_i^+)}}{e^{s(q_i, d_i^+)} + e^{s(q_i, d_i^-)} + \sum_j e^{s(q_i, d_{i,j}^-)}},
\end{equation}
where $s(q, d)$ denotes the ranking score obtained via dot product between query and document representations.

To encourage sparsity, SPLADE uses FLOPS regularization \cite{paria2020minimizingflopslearnefficient}, a smooth approximation of scoring operations tied to retrieval latency. For a batch of $\mathcal{N}$ documents, the average activation of token $j$ is estimated by $\bar{a}_j = \frac{1}{\mathcal{N}} \sum_{i=1}^\mathcal{N} w_j^{(d_i)}$, yielding:
\begin{equation}
    \mathcal{L}_\mathit{FLOPS} = \sum_{j \in V} \bar{a}_j^2 = \sum_{j \in V} \left(\frac{1}{\mathcal{N}} \sum_{i=1}^\mathcal{N} w_j^{(d_i)}\right)^2. 
\end{equation}
Unlike the $\ell_1$ regularization used in SNRM \cite{10.1145/3269206.3271800}, it penalizes high average term weights to promote a balanced index distribution. The overall objective is:
\begin{equation}
\label{eq:splade_loss}
    \mathcal{L} = \mathcal{L}_\mathit{rank\text{-}IBN} + \lambda_q \mathcal{L}_\mathit{FLOPS}^q + \lambda_d \mathcal{L}_\mathit{FLOPS}^d,
\end{equation}
where $\lambda_q$ and $\lambda_d$ control query and document sparsity, respectively. Later SPLADE variants use techniques such as knowledge distillation~\cite{10.1145/3477495.3531857,formalSPLADEV2Sparse2021}; we use standard contrastive learning to isolate our architectural and sparsification mechanisms, which are orthogonal to such enhancements.
\vspace{-2mm}
\subsection{Exploration and challenges} 
\label{subsec:Application and Challenges}
While SPLADE \cite{10.1145/3404835.3463098,formalSPLADEV2Sparse2021} has demonstrated a performance comparable to dense retrieval on several benchmarks, especially in passage retrieval, its practical effectiveness in product search scenarios remains underexplored. 

\heading{Difference between passage retrieval and product retrieval}
Passages are typically \emph{content-rich}, coherent natural language with many noisy terms, requiring models to identify important terms while filtering less informative content. Product search is more \emph{keyword-centric}. User queries are often non-fluent keyword aggregations that express purchase intent, while product titles stack attributes, brands, model identifiers, and other salient terms to maximize coverage of possible needs. Consequently, most query terms carry important constraints, and titles must support expansion without diluting these literal signals.

\heading{Implementing SPLADE in product search} On the Multi-CPR E-commerce test set \cite{10.1145/3477495.3531736}, our BERT-based SPLADE achieves 89.6\% Hit@1000, below DPR's 92.1\% and the 96\%+ reported on MS MARCO passage ranking \cite{bajaj2018msmarcohumangenerated}. 

\heading{Enhancing SPLADE with LLMs for product search}
Recent work has begun to adapt LLMs for sparse retrieval. \citet{zengScalingSparseDense2025} study the scaling behavior of LLM-based sparse retrievers, while CSPLADE \cite{xuCSPLADELearnedSparse2025}, drawing on NV-Embed \cite{leeNVEmbedImprovedTechniques2025} and Echo Embeddings \cite{springerRepetitionImprovesLanguage2024}, addresses the information-flow limitations of unidirectional attention in LLM-based SPLADE. Following this line of work, we replace BERT with Qwen2.5-3B \cite{qwen2025qwen25technicalreport}, increasing Hit@1000 to 91.3\% on Multi-CPR E-commerce \cite{10.1145/3477495.3531736}. Adapting existing strategies for improving LLM representation quality \cite{behnamghaderLLM2VecLargeLanguage2024,springerRepetitionImprovesLanguage2024,xuCSPLADELearnedSparse2025} further raises Hit@1000 to 92.4\%, but target-product ranking remains suboptimal.

\begin{table}[t]
\caption{Top-weighted terms from Qwen2.5-based SPLADE and PROSPER, in descending weight. Literal terms are black, useful expansions green italic, and noisy expansions red underlined.}
\label{tab:hallucination}
\centering
\includegraphics[width=\linewidth]{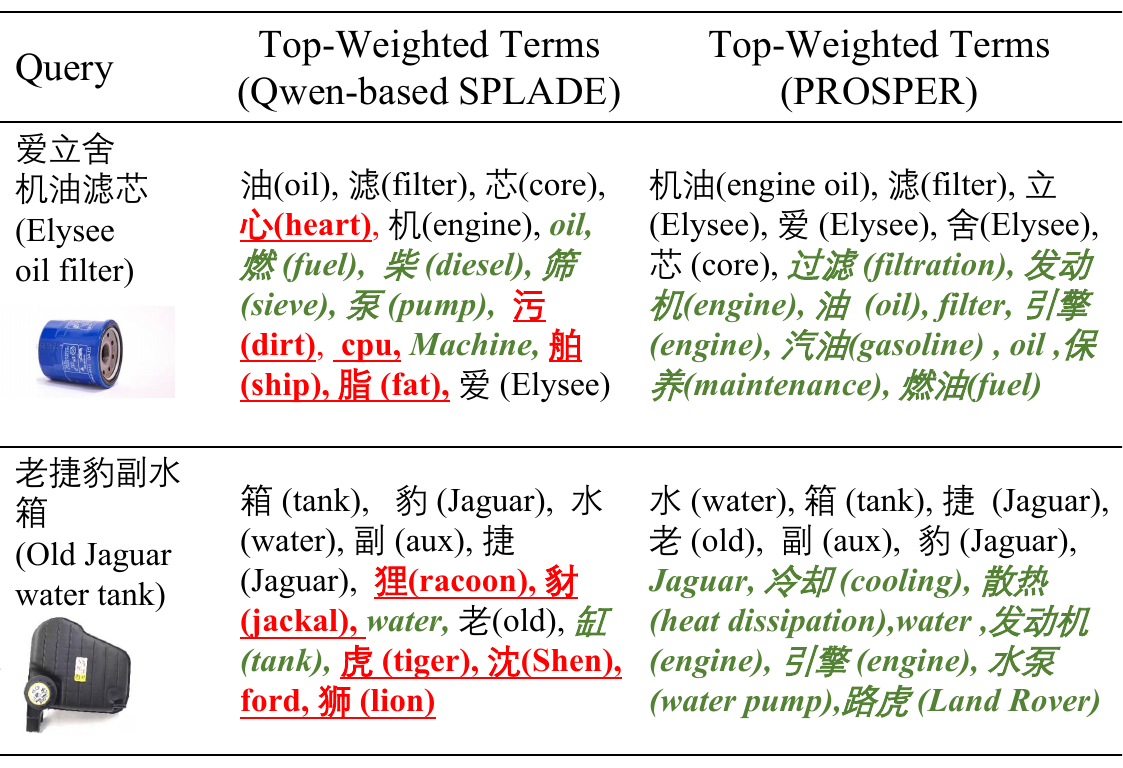}
\vspace{-3mm}
\end{table}

\heading{Case study and challenges}
Case analyses reveal severe \emph{lexical expansion hallucinations} (see Table~\ref{tab:hallucination}) in Qwen-based SPLADE:
\begin{enumerate*}[label=(\roman*)]
\item literal terms such as brands and product names remain underweighted
(e.g., \begin{CJK}{UTF8}{gbsn}“爱立舍机油滤芯”\end{CJK}, where ``Elysee'' was overlooked); and
\item irrelevant expansions appear in ambiguous or rare queries
(e.g., \begin{CJK}{UTF8}{gbsn}“老捷豹副水箱”\end{CJK}, where the model expanded irrelevant animal-related terms).
\end{enumerate*}

In product search, both the target product category and its modifiers are essential for capturing the user's actual intent.
If these literal terms are underweighted, the target product can be easily overwhelmed by a large number of superficially relevant products from the same category, leading the retriever to return items that appear semantically related but fail to match the user's specific need.

Moreover, LLM-based sparse retrieval also faces a training difficulty caused by the extremely large vocabulary space.
LLMs typically have very large vocabularies; for example, the Qwen2.5 model~\cite{qwen2025qwen25technicalreport} contains over 150,000 dimensions, which is excessive compared with the length of queries (average 6 terms in our experimental datasets) and items (average 25 terms in our experimental datasets).
Without proper constraints, sparsifying such high-dimensional representations leads to unstable early training and prevents the model from learning meaningful representations.

\label{sec:method}
\begin{figure*}[t]
    \centering
    \begin{subfigure}[b]{0.69\textwidth}
        \centering
        \includegraphics[width=\linewidth, trim={0 0 426.49pt 0}, clip]{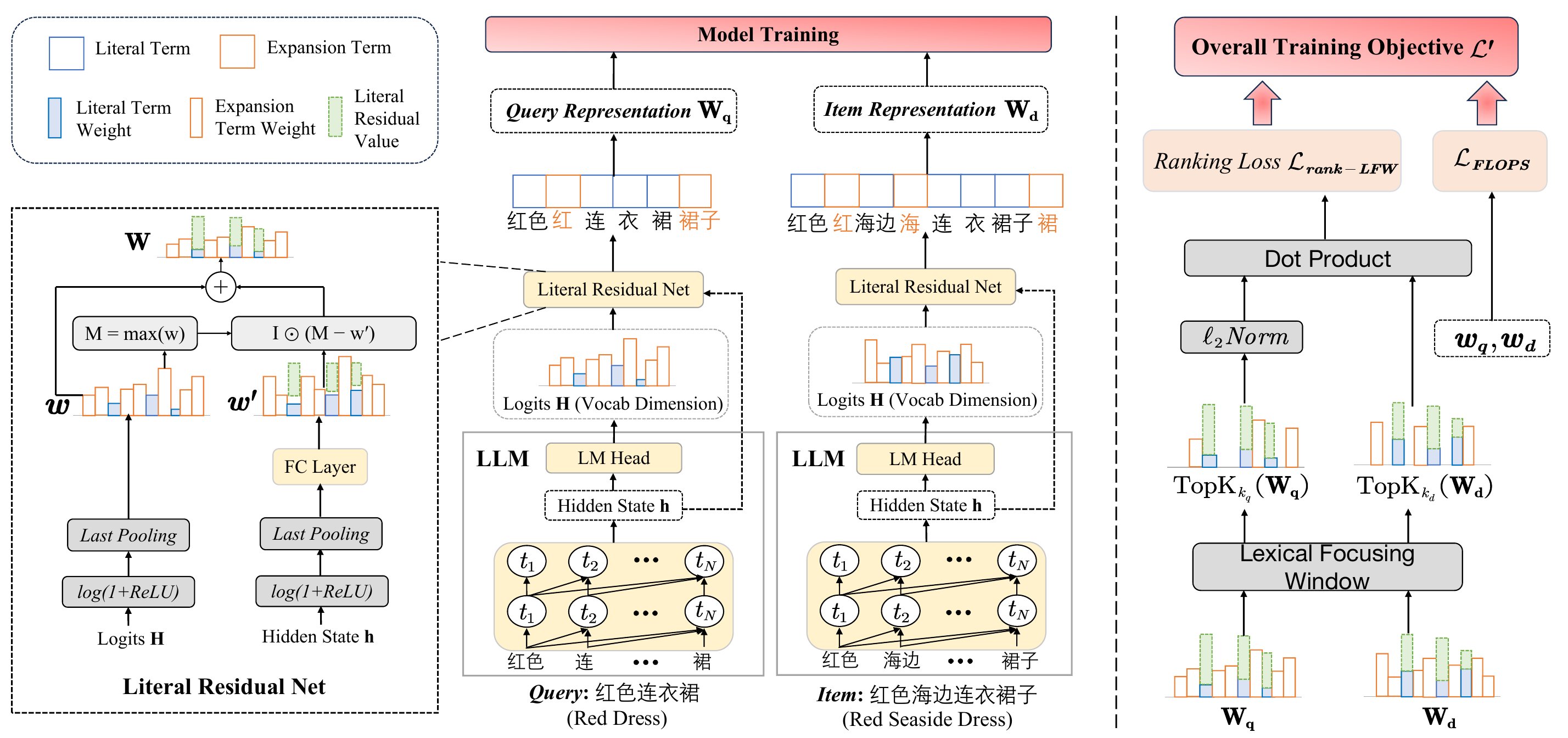}
        \caption{Model architecture}
        \label{fig:model_a}
    \end{subfigure}%
    \hfill
    \begin{subfigure}[b]{0.31\textwidth}
        \centering
        \includegraphics[width=\linewidth, trim={949.29pt 0 0 0}, clip]{figs/Overview.pdf}
        \caption{Model training}
        \label{fig:model_b}
    \end{subfigure}

    \caption{Overview of the proposed PROSPER framework. 
    (a) The model architecture, detailing the query/item encoding process and the core LRN module that calibrates literal term weights.
    (b) The model training, illustrating the LFW, the asymmetric similarity computation, and the final loss function combining ranking loss with FLOPS regularization.}
    \label{fig:overall_model}
\end{figure*}
\section{Method}

As illustrated in Figure~\ref{fig:overall_model}, PROSPER combines the literal residual network to alleviate lexical expansion hallucination with the lexical focusing window for targeted training initialization.

\subsection{Model architecture}
\label{sec:model_architecture}
\heading{LLM-based representation} 
Building upon SPLADE \cite{formalSPLADEV2Sparse2021}, our approach uses an LLM backbone to improve semantic understanding. For a tokenized query or item $S = (t_1, t_2, \ldots, t_N)$, we extract its last-layer hidden states $\mathbf{h} = (h_1, h_2, \ldots, h_N)$. The language modeling head projects each hidden state to vocabulary-sized logits $\mathbf{H}_i \in \mathbb{R}^{|V|}$, which, together with $\mathbf{h}$, are input to the literal residual network.

\heading{Literal residual network (LRN)}
The core idea of LRN is to guide the LLM to first focus on literal terms before expanding to related terms, assigning sufficient weights to critical terms such as product brands and product models that represent specific user needs.

For the hidden states $\mathbf{h}$ and logits $\mathbf{H}$ extracted from the LLM, we first apply ReLU activation and log-saturation transformation:
\begin{equation}
    \tilde{h}_{i} = \log(1 + \text{ReLU}(h_{i})), ~~~~~~
    \tilde{H}_{ij} = \log(1 + \text{ReLU}(H_{ij})),
\end{equation}
where $h_i$ is the hidden state vector for the $i$-th input token, and $H_{ij}$ represents the importance of the $j$-th vocabulary term for the $i$-th input token. 

While SPLADE uses max-pooling \cite{formalSPLADEV2Sparse2021}, causal LLMs aggregate the preceding context at the final token. We therefore adopt last-pooling \cite{taoLLMsAreAlso2024,behnamghaderLLM2VecLargeLanguage2024}, taking the representation of the last content token before the appended end-of-sequence token. The basic representation $\mathbf{w}$ is derived by last-pooling the transformed logits $\tilde{H}$:
\begin{equation}
    \mathbf{w}_{j} = \text{Last}(\tilde{H}_{ij}),
\end{equation}
where $\mathbf{w}_{j}$ is the weight for the $j$-th term in the vocabulary.

Meanwhile, we obtain a pooled hidden state representation by last-pooling the transformed hidden states $\tilde{h}$:
\begin{equation}
    \tilde{h}_{\text{last}} = \text{Last}(\tilde{h}).
\end{equation}
A fully connected layer projects $\tilde{h}_{\text{last}}$ to the vocabulary-sized enhancement vector $\mathbf{w}'\in\mathbb{R}^{|V|}$:
\begin{equation}
    \mathbf{w}' = \text{FC}_{\text{layer}}(\tilde{h}_{\text{last}}).
\end{equation}
The projection has no output activation. We combine the basic representation $\mathbf{w}$ and the learned vector $\mathbf{w}'$ to obtain the final representation $\mathbf{W}$:
\begin{equation}
    \mathbf{W} = \mathbf{w} + \mathbf{I} \odot (\max(\mathbf{w}) - \mathbf{w}'),
\end{equation}
where $\mathbf{I}$ is 1 at literal-term positions in $S$ and 0 elsewhere, $\odot$ denotes element-wise multiplication, and $\max(\mathbf{w})$ is the largest weight in the basic representation. The learned residual $\max(\mathbf{w})-w'_j$ adjusts each literal term, while expansion weights remain unchanged. We use both $\mathbf{W}$ and $\mathbf{w}$ to train the model.

\heading{Discussion}
LRN acts as a compensatory weighting mechanism. The pretrained LM head produces $\mathbf{w}$ for semantic expansion, while a trainable projection produces $\mathbf{w}'$ for literal-term calibration. The maximum $\max(\mathbf{w})$ provides an input-dependent reference for the strongest activation that can overshadow a weak literal term. Subtracting $w'_j$ from this reference gives the learned compensation for literal term $j$: a lower $w'_j$ yields more compensation, and a higher value yields less. Adding this residual to $w_j$ retains its base contribution while allowing term-specific adjustment. Retrieval supervision learns how much compensation each literal term needs.

We also explored two alternative compensation strategies.
\begin{enumerate*}[label=(\roman*)]
\item \emph{Self-compensation without $\mathbf{w}'$}, which directly compensates literal terms using the basic representation:
$
\mathbf{W} = \mathbf{w} + \mathbf{I} \odot (\max(\mathbf{w}) - \mathbf{w}).
$
This variant, denoted as $\mathrm{PROSPER}_{\text{LRN-selfcomp}}$, sets every literal weight to $\max(\mathbf{w})$, removing their weight differences;
\item \emph{Direct addition without complementary compensation}, which directly adds the enhancement vector:
$
\mathbf{W} = \mathbf{w} + \mathbf{I} \odot \mathbf{w}'.
$
This variant, denoted as $\mathrm{PROSPER}_{\text{LRN-add}}$, learns the adjustment entirely through $\mathbf{w}'$. LRN supplies the input-dependent reference $\max(\mathbf{w})$ explicitly, helping weak literals compete with strong expansions across inputs with different activation scales.
\end{enumerate*}
In the LRN ablation study (Section~\ref{sec:lrn-ablation}), both alternatives perform worse than the complementary residual design on our benchmark.

\subsection{Model training}
\label{sec:model_training}
\heading{Coarse-to-fine sparsification}
FLOPS regularization controls sparsity softly. Our coarse-to-fine strategy enforces rapid sparsification in early training to establish a stable initialization, then refines term weights to balance retrieval quality and sparsity.

\heading{Lexical focusing window}
Top-$k$ sparsification has been studied in SPLADE-mask~\cite{yang2021topk} and AdaSparse~\cite{xie2026adasparse}. We adopt it as a \emph{lexical focusing window} (LFW) for product search, where it improves retrieval and accelerates sparsification. LFW applies a conditional top-$k$ pooling operator $\text{TopK}_k$ that activates only when the number of non-zero dimensions in a representation $\mathbf{W}$ exceeds the window size $k$:
\begin{equation}
\label{eq:conditional_topk}
\text{TopK}_k(\mathbf{W}) = 
\begin{cases} 
    \text{TopK}(\mathbf{W}, k) & \text{if } ||\mathbf{W}||_0 > k, \\
    \mathbf{W} & \text{if } ||\mathbf{W}||_0 \le k, 
\end{cases}
\end{equation}
where $||\cdot||_0$ is the $\ell_0$ norm, counting the non-zero elements in the vector, and the standard $\text{TopK}$ function retains the $k$ largest weights while setting others to zero. Thus, LFW primarily affects the early, dense stages and phases out as representations become naturally sparse.

LFW focuses early ranking supervision on the most important terms, enabling rapid sparsification. As fewer dimensions remain active, its role diminishes. FLOPS regularization operates throughout training and continues to refine the sparse weights.

\heading{Training objective}
Our final objective follows SPLADE's contrastive ranking loss. Unlike its standard dot product, we apply $\ell_2$ normalization only to the query representation, which improves sparsification and retrieval performance in our experiments (Section~\ref{norm}). 
The similarity score is computed using the LFW operator:
\begin{equation}
\label{eq:tbsplade_similarity}
    s_{LFW}(\mathbf{W}_q, \mathbf{W}_d) = \frac{\text{TopK}_{k_q}(\mathbf{W}_q)}{||\text{TopK}_{k_q}(\mathbf{W}_q)||_2} \cdot \text{TopK}_{k_d}(\mathbf{W}_d),
\end{equation}
where $\mathbf{W}_q$ and $\mathbf{W}_d$ are the query and item representations from the LRN output, $k_q$ and $k_d$ are their respective LFW sizes, and $||\cdot||_2$ is the $\ell_2$ norm. 
Dynamic window shrinking did not improve over the best fixed setting in our experiments, suggesting that LFW's main benefit lies in providing a robust and efficient training start. 

We use InfoNCE \cite{oord2019representationlearningcontrastivepredictive} with in-batch negatives \cite{lin-etal-2021-batch}. For query $q_i$, positive item $d_i^+$, and negatives $\{d_{i,j}^-\}_j$ from other queries in the batch, the ranking loss is:
\begin{equation}
\label{eq:infonce_loss}
    \mathcal{L}_\mathit{rank\text{-}LFW} = -\log \frac{e^{s_{LFW}(\mathbf{W}_{q_i}, \mathbf{W}_{d_i^+})}}{e^{s_{LFW}(\mathbf{W}_{q_i}, \mathbf{W}_{d_i^+})} + \sum_j e^{s_{LFW}(\mathbf{W}_{q_i}, \mathbf{W}_{d_{i,j}^-})}}.
\end{equation}
The overall training objective combines this ranking loss with the FLOPS regularization terms \cite{paria2020minimizingflopslearnefficient} for both queries and items:
\begin{equation}
\label{eq:tbsplade_loss}
    \mathcal{L}' = \mathcal{L}_\mathit{rank\text{-}LFW} + \lambda_q \mathcal{L}_{\mathit{FLOPS}}^q + \lambda_d \mathcal{L}_{\mathit{FLOPS}}^d,
\end{equation}
where $\lambda_q$ and $\lambda_d$ are hyperparameters that balance the regularization of FLOPS. Here, FLOPS regularization is applied to the basic representations $\mathbf{w}_q$ and $\mathbf{w}_d$.

\label{-2mm}
\section{Offline Experiments}
\label{sec:experiment}
In this section, we conduct a series of offline experiments in a product search scenario to comprehensively evaluate PROSPER.

\label{-1mm}
\subsection{Experimental setup}
\textbf{Datasets.} We conduct experiments on two datasets that represent public benchmarks and real-world industrial settings:
\begin{enumerate*}[label=(\roman*)]
\item \textbf{Multi-CPR e-commerce} \cite{10.1145/3477495.3531736}. Multi-CPR is a publicly available, multi-domain Chinese passage retrieval dataset. For our experiments, we use its e-commerce subset, which is sourced from real-world search scenarios. This subset contains a corpus of over 1 million passages, from which we use a training set of 100,000 query-passage pairs and a test set of 1,000 queries. 
\item \textbf{Real-world e-commerce}. To further validate our approach in an industrial setting, we construct a new dataset by sampling approximately 1.07 million query-item pairs from the real user click logs of a real-world e-commerce platform. The user clicks serve as the ground truth for relevance. A single query in this dataset may correspond to 1 to 10 clicked items. We create a training set of 270,000 query-item pairs and a test set of 1,000 queries along with their associated clicked items. 
\end{enumerate*}
The query and item length statistics after tokenization by the Qwen2.5 tokenizer are provided in Table~\ref{tab:dataset_stats_full}.

\begin{table}[t]
\centering
\caption{Full token length statistics of the datasets.}
\label{tab:dataset_stats_full}
\begin{tabular}{@{}l ccc @{\quad} ccc@{}}
\toprule
\multirow{2}{*}{\textbf{Dataset}} &
\multicolumn{3}{c}{\textbf{Query length}} &
\multicolumn{3}{c}{\textbf{Item length}} \\
\cmidrule{2-4} \cmidrule{5-7}
 & Min & Avg & Max & Min & Avg & Max \\
\midrule
Multi-CPR E-commerce & 1 & 5 & 23 & 2 & 25 & 92 \\
Real-world E-commerce & 2 & 6 & 25 & 2 & 25 & 74 \\
\bottomrule
\end{tabular}
\end{table}

\heading{Baselines}
For dense baselines, we compare against: 
\begin{enumerate*}[label=(\roman*)]
\item \textbf{DPR} \cite{karpukhin-etal-2020-dense}, classic dense retrieval baseline.
\item \textbf{BGE Series} \cite{bge_embedding}, a series of powerful dense text embedding models, and we use \textbf{bge-large-zh-v1.5} and \textbf{bge-base-zh-v1.5} for comparison.
\end{enumerate*} 
For sparse baselines, we use: 
\begin{enumerate*}[label=(\roman*)]
\item \textbf{BM25} \cite{10.1561/1500000019}, the classic sparse method. We compare with \textbf{BM25}\textsubscript{BERT} and \textbf{BM25}\textsubscript{Qwen}, which use tokenizers of BERT-base-chinese and Qwen2.5-3B, respectively. 
\item \textbf{Doc2Query} \cite{nogueira2019documentexpansionqueryprediction}, which alleviates vocabulary mismatch by using a seq2seq model to generate potential queries for a document. 
\item \textbf{DeepCT}
\cite{dai2019contextawaresentencepassagetermimportance,10.1145/3397271.3401204}, which uses BERT to evaluate literal term importance. 
\item \textbf{SPLADE} \cite{10.1145/3404835.3463098,formalSPLADEV2Sparse2021}, our main baseline. We use both \textbf{SPLADE} \cite{10.1145/3404835.3463098} and \textbf{SPLADE-v2} \cite{formalSPLADEV2Sparse2021} and adjust their training process for a fair comparison.
\item We implement \textbf{SP}\textsubscript{Qwen-backbone}~\cite{zengScalingSparseDense2025} by replacing the BERT backbone in SPLADE with Qwen2.5-3B to examine whether simply scaling sparse retrieval to an LLM backbone is sufficient for product search. Following recent work on causal LLM-based sparse retrieval, particularly
CSPLADE \cite{xuCSPLADELearnedSparse2025}, and representative LLM embedding
methods for improving representation quality, including Echo Embeddings
\cite{springerRepetitionImprovesLanguage2024}, LLM2Vec
\cite{behnamghaderLLM2VecLargeLanguage2024}, and NV-Embed
\cite{leeNVEmbedImprovedTechniques2025}, we construct two strengthened
Qwen-based sparse baselines. \textbf{SP}\textsubscript{Qwen-echoembedding}
adapts the input-repetition strategy of Echo Embeddings, while
\textbf{SP}\textsubscript{Qwen-bidattention} enables bidirectional attention.
\end{enumerate*}

\heading{Model variants}
To validate the effectiveness of our proposed components, we create several variants of PROSPER.

\begin{itemize}[leftmargin=*,nosep]
\item To analyze the core model design, the variants are: 
\begin{enumerate*}[label=(\roman*)]
\item \textbf{PROSPER}\textsubscript{BERT}, which replaces the Qwen2.5-3B \cite{qwen2025qwen25technicalreport} backbone with BERT-base-chinese \cite{devlin-etal-2019-bert};
\item \textbf{PROSPER}\textsubscript{max-pooling}, which uses the max-pooling strategy from SPLADE;
\item \textbf{PROSPER}\textsubscript{bid-attention}, which uses bidirectional instead of causal attention \cite{behnamghaderLLM2VecLargeLanguage2024,xuCSPLADELearnedSparse2025};
and \item \textbf{PROSPER}\textsubscript{echo-emb}, which duplicates the input sequence to simulate a bidirectional receptive field \cite{springerRepetitionImprovesLanguage2024,xuCSPLADELearnedSparse2025,leeNVEmbedImprovedTechniques2025}.
\end{enumerate*}

\item To investigate the role of the LRN, the variants are: 
\begin{enumerate*}[label=(\roman*)]
\item \textbf{PROSPER}\textsubscript{w/o-LRN}, which removes the LRN module entirely;
\item \textbf{PROSPER}\textsubscript{LRN-selfcomp}, which does not use the learned enhancement vector $\mathbf{w}'$ and directly compensates literal terms based on the basic representation
($\mathbf{W} = \mathbf{w} + \mathbf{I} \odot (\max(\mathbf{w}) - \mathbf{w})$);
\item \textbf{PROSPER}\textsubscript{LRN-add}, which replaces the complementary residual with direct addition
($\mathbf{W} = \mathbf{w} + \mathbf{I} \odot \mathbf{w}'$);
and \item variants that apply LRN only to queries (\textbf{PROSPER}\textsubscript{LRN-q}) or items (\textbf{PROSPER}\textsubscript{LRN-d}).
\end{enumerate*}

\item To evaluate the LFW, the variants are: 
\begin{enumerate*}[label=(\roman*)]
\item \textbf{PROSPER}\textsubscript{w/o-LFW}, which removes LFW; 
\item \textbf{PROSPER}\textsubscript{LFW-dynamic}, which uses a dynamic window size;
and \item using various fixed window sizes for query ($k_q$) and item ($k_d$).
\end{enumerate*}

\item To study the literal and expansion terms, we test: 
\begin{enumerate*}[label=(\roman*)]
\item \textbf{PROSPER}\textsubscript{literal} and \textbf{PROSPER}\textsubscript{expansion}, which are trained and evaluated using only literal or expansion terms, respectively;
and \item \textbf{PROSPER}\textsubscript{mask-lit} and \textbf{PROSPER}\textsubscript{mask-expan}, where the model is trained normally, but either literal or expansion terms are masked out during match and evaluation stage.
\end{enumerate*}

\item To examine query--item asymmetry in normalization, we test:
\begin{enumerate*}[label=(\roman*)]
\item \textbf{PROSPER}\textsubscript{all-norm}, which applies $\ell_2$ normalization to both query and item vectors, making the scoring function equivalent to standard cosine similarity;
\item \textbf{PROSPER}\textsubscript{w/o-norm}, which removes all normalization, resulting in a standard dot product score;
\item \textbf{PROSPER}\textsubscript{d-norm}, which applies $\ell_2$ normalization only to the item vectors;
and \item \textbf{PROSPER}\textsubscript{$\ell_1$-norm}, which applies $\ell_1$ normalization to the query vector to explore its effect on term weighting.
\end{enumerate*}

\end{itemize}

\begin{table*}[t]
\centering
\caption{Main results (\%) on the Multi-CPR E-commerce and Real-world
E-commerce datasets. Best results are in bold. “–” denotes metrics not reported in the original Multi-CPR paper \cite{10.1145/3477495.3531736}.}
\label{tab:main_results}
\setlength{\tabcolsep}{4.0pt}
\begin{tabular}{l l ccccc c ccc}
\toprule
& \multirow{2}{*}{\textbf{Method}} & \multicolumn{5}{c}{\textbf{Multi-CPR E-commerce}} & & \multicolumn{3}{c}{\textbf{Real-world
E-commerce}} \\
\cmidrule(lr){3-7} \cmidrule(lr){9-11}
& & Hit@1 & Hit@10 & Hit@100 & Hit@1000 & MRR@10 & & Recall@10 & Recall@100 & Recall@1000 \\
\midrule
\multirow{3}{*}[-1pt]{\textit{Dense Baselines}}
& DPR & - & - & - & 92.1 & 27.04 & & 42.50 & 72.61 & 91.07 \\
& BGE-base-zh-v1.5 & \textbf{26.6} & 54.1 & 80.4 & 93.7 & 34.70 & & 50.83 & 75.87 & 92.93 \\
& BGE-large-zh-v1.5 & 26.1 & \textbf{55.2} & \textbf{81.9} & 93.1 & \textbf{35.10} & & 50.66 & 76.89 & 93.45 \\
\midrule
\multirow{9}{*}[-1pt]{\textit{Sparse Baselines}}
& BM25\textsubscript{Qwen} & 14.3 & 37.1 & 62.6 & 83.7 & 20.60 & & 39.80 & 65.53 & 84.68 \\
& BM25\textsubscript{BERT} & 16.5 & 40.9 & 66.9 & 85.5 & 23.48 & & 41.23 & 67.84 & 86.21 \\
& Doc2Query & - & - & - & 82.6 & 23.85 & & 42.06 & 68.41 & 86.72 \\
& DeepCT & 21.9 & 46.4 & 72.2 & 87.5 & 29.14 & & 46.64 & 73.26 & 86.90 \\
& SPLADE & 17.2 & 42.7 & 70.6 & 89.2 & 25.87 & & 44.41 & 72.67 & 89.96 \\
& SPLADE-v2 & 18.4 & 44.1 & 71.3 & 89.6 & 26.40 & & 45.28 & 73.41 & 90.38 \\
& SP\textsubscript{Qwen-backbone} & 18.2 & 43.7 & 72.3 & 91.3 & 25.46 & & 45.12 & 73.46 & 91.79 \\
& SP\textsubscript{Qwen-echoembedding} & 19.7 & 44.0 & 74.2 & 92.4 & 26.72 & & 45.27 & 74.35 & 92.84 \\
& SP\textsubscript{Qwen-bidattention} & 19.5 & 43.8 & 73.5 & 91.8 & 26.41 & & 45.39 & 73.89 & 92.10 \\
\midrule
\multirow{5}{*}[-1pt]{\textit{Ours}}
& PROSPER & 25.3 & 50.7 & 78.1 & 93.9 & 32.85 & & 50.90 & 76.20 & 94.08 \\
& PROSPER\textsubscript{BERT} & 23.6 & 49.2 & 75.8 & 91.5 & 31.83 & & 50.43 & 75.97 & 93.68 \\
& PROSPER\textsubscript{max-pooling} & 22.6 & 49.5 & 78.3 & 93.6 & 30.85 & & 48.45 & 76.33 & 93.75 \\
& PROSPER\textsubscript{echo-emb} & 24.5 & 50.7 & 80.3 & \textbf{94.1} & 32.82 & & 50.18 & \textbf{77.20} & \textbf{94.43} \\
& PROSPER\textsubscript{bid-attention} & 22.3 & 51.4 & 78.7 & 93.6 & 30.72 & & \textbf{51.06} & 76.51 & 93.80 \\
\bottomrule
\end{tabular}
\vspace*{-2mm}
\end{table*}

\vspace{-2mm}
\subsection{Experimental results}
\textbf{Overall performance.}
Table~\ref{tab:main_results} compares PROSPER with various baselines on both datasets and highlights three main findings.
\begin{itemize}[leftmargin=*,nosep]

\item \emph{Compared with sparse baselines}, PROSPER achieves the best performance across all metrics on both datasets. PROSPER\textsubscript{BERT} also outperforms SPLADE-v2, showing that the benefits of our design are not limited to the Qwen2.5 backbone. Together, these results suggest that preserving literal information while learning sparse representations is beneficial across the two backbone architectures evaluated in this study.

\item \emph{Compared with dense baselines}, PROSPER remains competitive, particularly on Hit@1000. The BGE models perform better on Hit@1 and MRR@10, demonstrating their strength at shallow retrieval cutoffs. In addition to the results reported in Table~\ref{tab:main_results}, we explored several other dense retrieval settings on Multi-CPR E-commerce:
\begin{enumerate*}[label=(\roman*)]
\item fine-tuning the BGE models with standard contrastive loss did not improve their performance and instead led to degradation;
\item fine-tuning Qwen2.5-3B as a dense retriever with contrastive loss also yielded limited effectiveness; and
\item the strongest dense model we tested, Qwen3-Embedding-4B~\cite{zhang2025qwen3embedding}, achieved a Hit@1000 of 94.9.
\end{enumerate*}
PROSPER provides an interpretable, complementary retrieval signal that brings incremental value to online multi-channel recall systems.

\item \emph{Compared with model variants}, the performance difference between PROSPER and PROSPER\textsubscript{BERT} suggests that the stronger semantic capacity of Qwen2.5 benefits the proposed framework. PROSPER also outperforms PROSPER\textsubscript{max-pooling}, particularly on MRR@10, supporting the use of last-token pooling in our setting. PROSPER\textsubscript{echo-emb} yields modest improvements on several metrics, including Hit@1000, but these gains come at the cost of doubling the input sequence length. PROSPER\textsubscript{bid-attention} shows no consistent improvement over the causal-attention model.
\end{itemize}

\begin{figure}[t]
\centering
\includegraphics[width=\columnwidth]{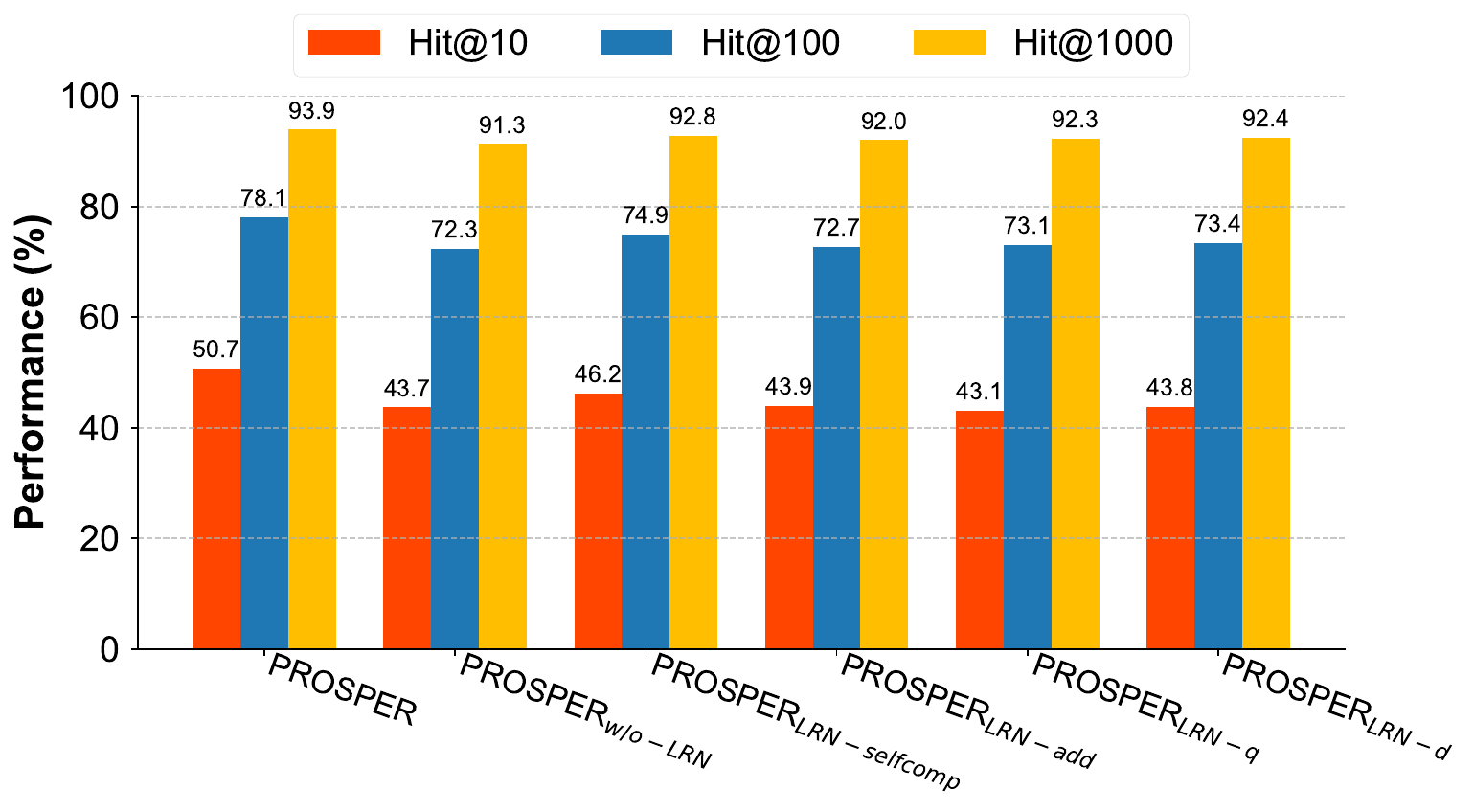}
\caption{Impact of the LRN on retrieval performance.}
\label{fig:lrn_impact}
\end{figure}

\heading{Impact of LRN} 
\label{sec:lrn-ablation}
Removing LRN reduces performance across all metrics (Figure~\ref{fig:lrn_impact}). Applying LRN only to queries or only to items gives similar improvements over this baseline, showing that LRN benefits both sides. Both alternative LRN variants perform worse than PROSPER. Self-compensation removes differences among literal weights, while direct addition learns the adjustment without a shared scale reference. LRN uses the base maximum to guide learned compensation for each literal term, helping retain terms with low initial weights.

\begin{table}[b]
\centering
\caption{Impact of the LFW on retrieval performance (\%).}
\label{tab:lfw_impact}
\setlength{\tabcolsep}{3pt}
\begin{tabular}{@{}l@{\hspace{4pt}}c@{\hspace{4pt}}c@{\hspace{4pt}}c@{\hspace{4pt}}c@{}}
\toprule
\textbf{Variant} & \textbf{Hit@1} & \textbf{Hit@10} & \textbf{Hit@100} & \textbf{Hit@1000} \\
\midrule
$k_d$=1024, $k_q$=512 & 22.3 & 49.2 & 77.7 & 93.5 \\
$k_d$=512, $k_q$=256 & 25.3 & 50.7 & 78.1 & 93.9 \\
$k_d$=256, $k_q$=128 & 23.7 & 50.2 & 77.0 & 93.4 \\
$k_d$=128, $k_q$=64 & 23.5 & 49.1 & 77.2 & 92.9 \\
$k_d$=64, $k_q$=32 & 21.9 & 48.3 & 74.3 & 91.9 \\
PROSPER\textsubscript{LFW-dynamic} & 24.6 & 49.0 & 77.8 & 93.4 \\
PROSPER\textsubscript{w/o-LFW} & 13.3 & 36.2 & 65.6 & 87.2 \\
\bottomrule
\end{tabular}
\end{table}

\heading{Impact of LFW}
Table~\ref{tab:lfw_impact} compares different LFW sizes and shows that:
\begin{enumerate*}[label=(\roman*)]
\item Without LFW (PROSPER\textsubscript{w/o-LFW}), performance drops across all metrics, showing the benefit of focusing training on the top relevant terms;
\item Among the different window sizes, the setting $k_q$ = 256, $k_d$ = 512 achieves the best overall performance, striking an optimal balance between focusing on the most relevant terms and maintaining sufficient vocabulary coverage;
\item PROSPER\textsubscript{LFW-dynamic}, with shrinking windows $(256, 128, 64)$ for queries and $(512, 256, 128)$ for items, performs slightly below the best fixed configuration. A well-chosen fixed window is sufficient for effective training.
\end{enumerate*}

\begin{figure}[t]
    \centering
    \includegraphics[width=\linewidth]{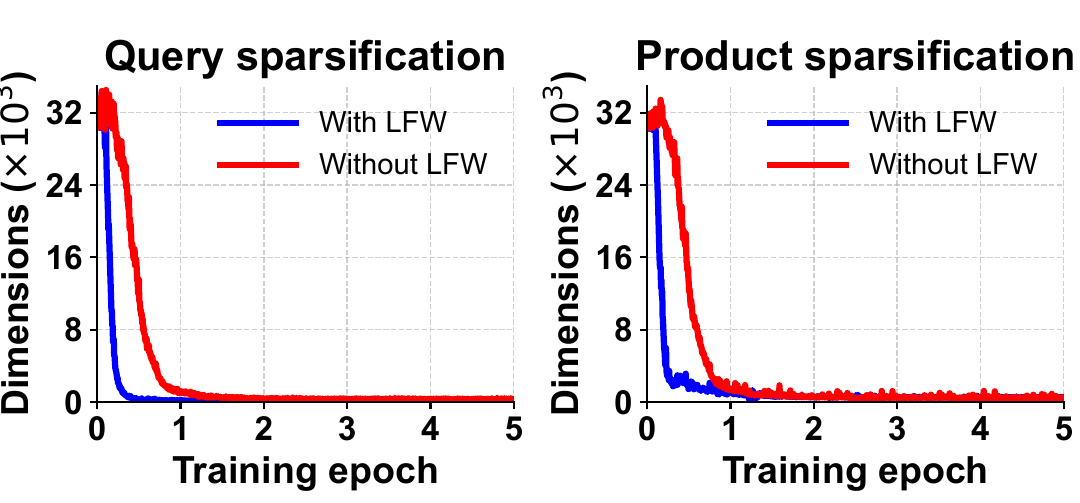}
    \caption{Evolution of active query and item dimensions during training with and without LFW.}
    \label{fig:sparsity_comparison}
\end{figure}

Moreover, we explore the sparsity evolution of query and item representations during training with and without LFW in Figure~\ref{fig:sparsity_comparison} and find that LFW enables rapid dimension reduction in early training stages while achieving faster stabilization. These results show that top-$k$ focusing is effective in product search: it accelerates convergence to sparse representations, while FLOPS regularization refines their weights throughout training.

\begin{figure}[htbp]
\centering
\includegraphics[width=\linewidth]{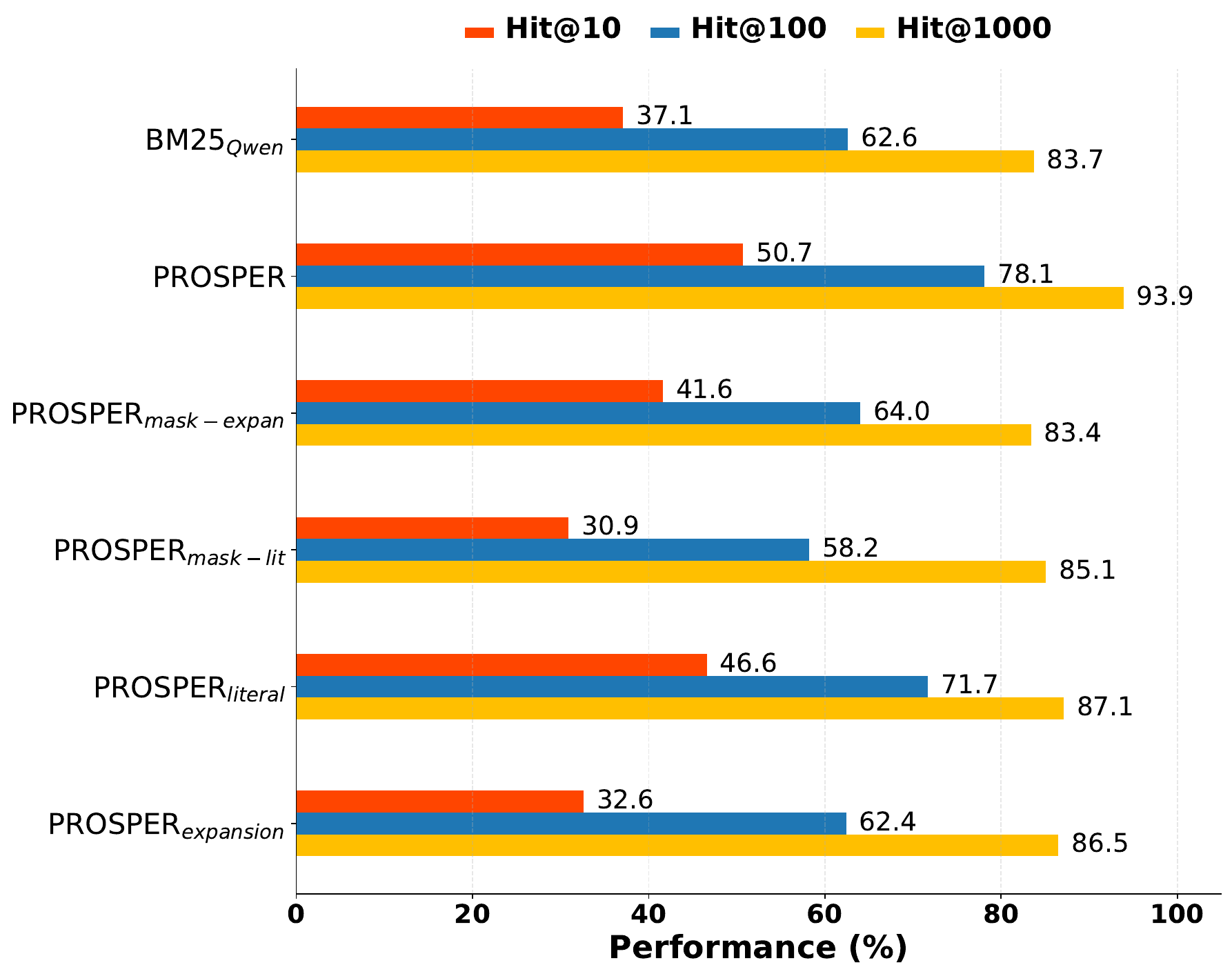}
\caption{Analysis of the impact of literal terms and expansion terms on retrieval performance.}
\label{fig:literal_vs_expansion}
\end{figure}

\noindent\textbf{Literal terms vs.\ expansion terms.}
To analyze the contributions of literal and expansion terms, we compare several variants in Figure~\ref{fig:literal_vs_expansion} and find:
\begin{enumerate*}[label=(\roman*)]
\item The PROSPER\textsubscript{literal} model, trained only on literal terms, outperforms the strong BM25 baseline (e.g., +9.5 points in Hit@10), demonstrating the effectiveness of our model in evaluating the importance of terms even without expansion;
\item Literal terms are superior for precision. This is evident as PROSPER\textsubscript{literal} and PROSPER\textsubscript{mask-expan} outperform their expansion-only counterpart, PROSPER\textsubscript{expansion} and PROSPER\textsubscript{mask-lit}, in metrics such as Hit@10 and Hit@100;
\item Expansion terms are crucial for supplementing recall. When trained in isolation, the recall of PROSPER\textsubscript{expansion} nearly matches that of PROSPER\textsubscript{literal} at Hit@1000. More compellingly, when evaluating the fully trained model by masking components, the expansion-only variant (PROSPER\textsubscript{mask-lit}) even surpasses the literal-only variant (PROSPER\textsubscript{mask-expan}) on the Hit@1000.
\end{enumerate*}

Literal terms support precision, while expansions supplement recall. PROSPER combines their strengths to improve overall performance. 

\begin{figure}[t]
    \centering
    \includegraphics[width=\linewidth]{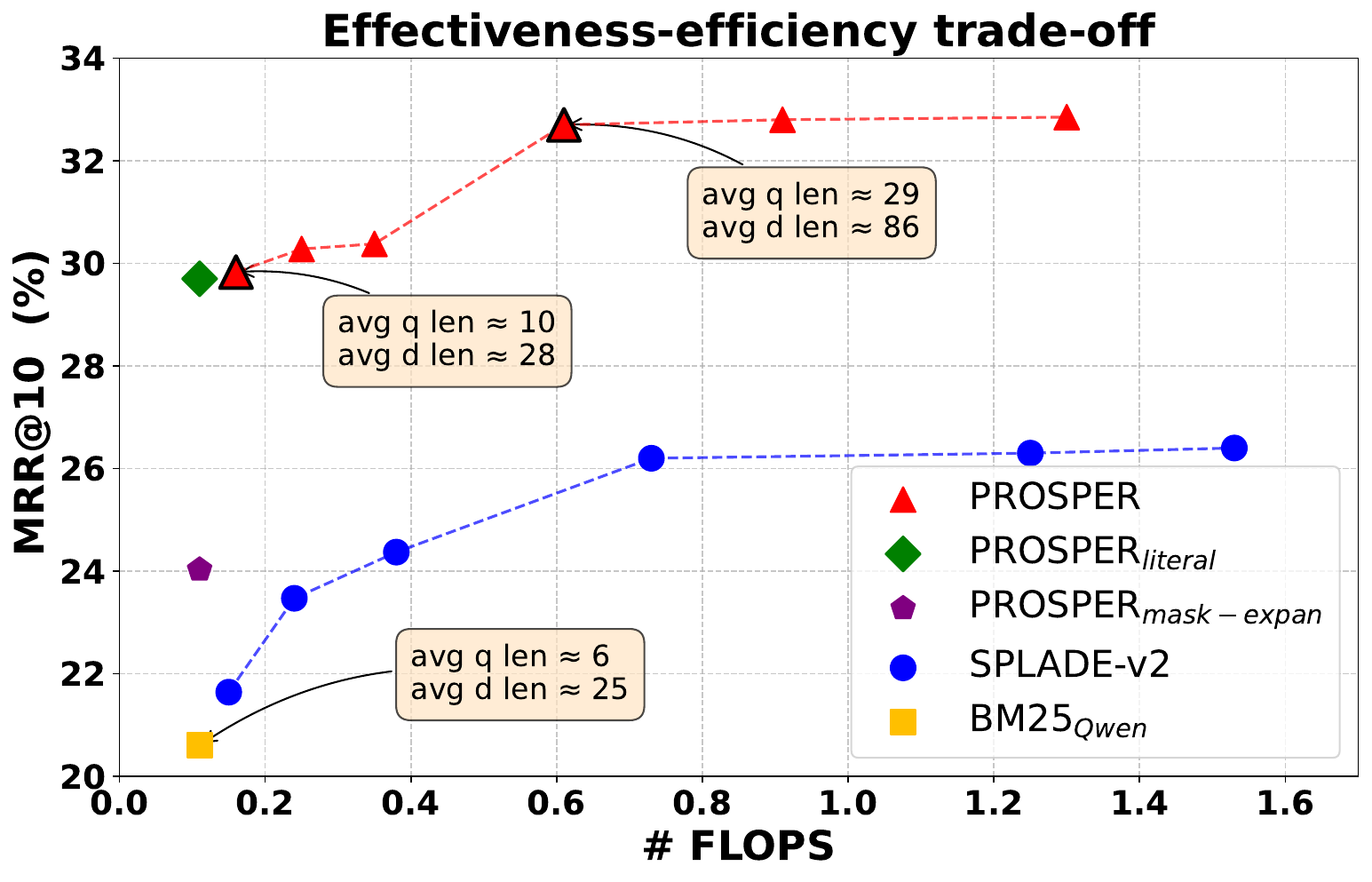}
    \caption{Effectiveness (MRR@10) vs. Efficiency (\# FLOPS) trade-off on Multi-CPR E-commerce. }
    \label{fig:efficiency_tradeoff}
\end{figure}

\heading{Effectiveness-efficiency trade-off}
We plot MRR@10 against \# FLOPS, the average number of overlapping query and item terms \cite{10.1145/3404835.3463098}. We control this matching cost through the FLOPS penalty and LFW size. Figure~\ref{fig:efficiency_tradeoff} shows that:
\begin{enumerate*}[label=(\roman*)]
\item Both PROSPER and SPLADE can achieve strong performance at low computational costs, but PROSPER consistently demonstrates a superior trade-off;
\item Compared to BM25, PROSPER\textsubscript{literal} and PROSPER\textsubscript{mask-expan} substantially improve retrieval quality at comparable matching cost. 
\end{enumerate*}

\begin{figure}[t]
    \centering
    \includegraphics[width=\linewidth]{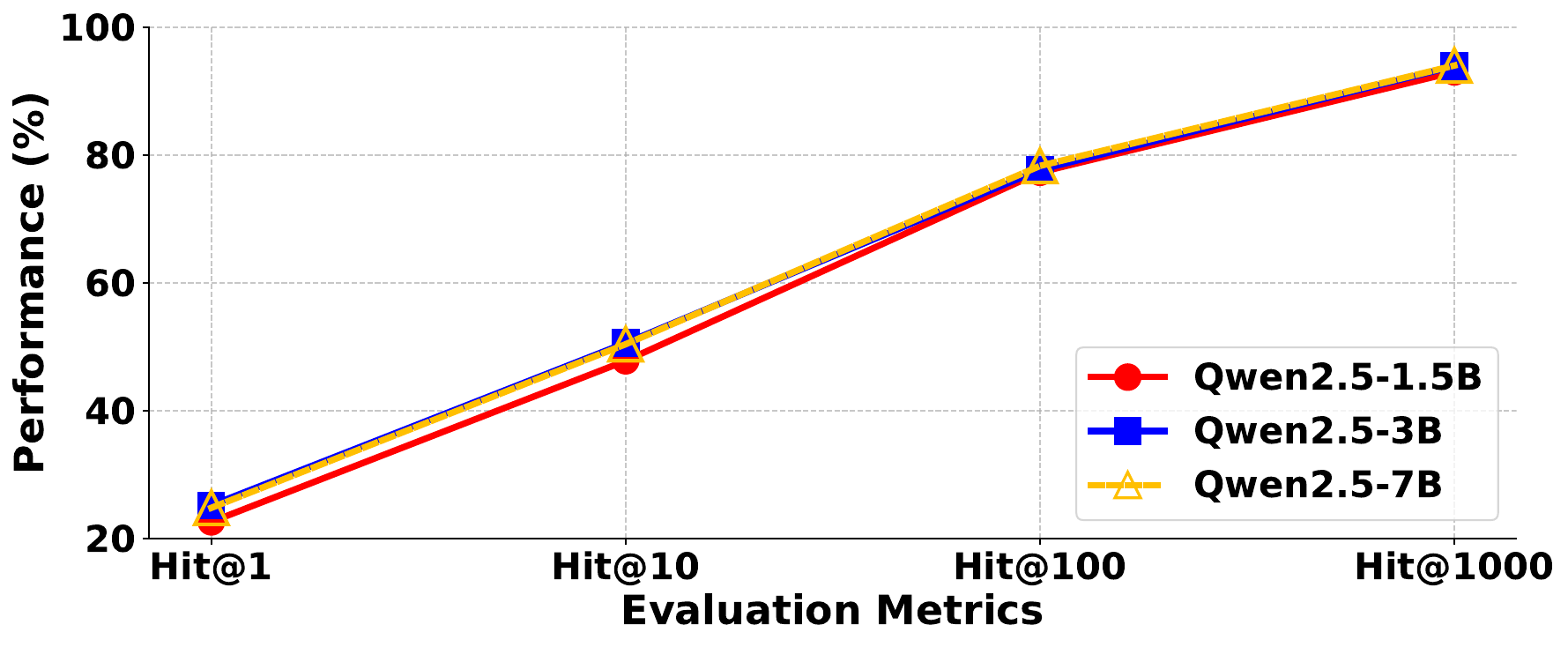}
    \caption{Performance comparison of PROSPER with different model sizes on the Multi-CPR test set.}
    \label{fig:parameter_scaling}
\end{figure}

We investigate model size scaling in Figure~\ref{fig:parameter_scaling} and find no significant performance gains with larger models. Therefore, we use the Qwen2.5-1.5B model as our backbone for online testing to achieve an optimal trade-off between effectiveness and cost.

\heading{Impact of normalization} \label{norm}
Table~\ref{tab:norm_impact} shows that query-only $\ell_2$ normalization performs best. During training, normalization promotes sparsification through competition for a shared norm. General query terms consume this norm while contributing little to distinguishing relevant products. Query-side $\ell_2$ normalization encourages the model to suppress these terms and emphasize useful, discriminative terms. Query-side $\ell_1$ normalization produces overly sparse queries in our experiments, suppressing useful terms and reducing recall.

Item attributes serve different queries. Normalizing an item makes unmatched attributes compete with useful matching terms for the shared norm, weakening their contribution. Applying $\ell_2$ normalization to both sides also over-sparsifies the representations in our experiments, reducing Hit@1000 to 87.2\%. Query-only $\ell_2$ normalization supports sparsification and discriminative weighting while preserving item attribute coverage.

\begin{table}[t]
\centering
\caption{Impact of different similarity functions on retrieval performance.}
\label{tab:norm_impact}
\setlength{\tabcolsep}{6pt}
\begin{tabular}{@{}l@{\hspace{6pt}}c@{\hspace{6pt}}c@{\hspace{6pt}}c@{\hspace{6pt}}c@{}}
\toprule
\textbf{Variant} & \textbf{Hit@1} & \textbf{Hit@10} & \textbf{Hit@100} & \textbf{Hit@1000} \\
\midrule
PROSPER & 25.3 & 50.7 & 78.1 & 93.9 \\
PROSPER\textsubscript{$\ell_1$-norm}   & 15.8 & 39.9 & 68.0 & 90.0 \\
PROSPER\textsubscript{w/o-norm}   & 17.0 & 40.9 & 68.0 & 90.0 \\
PROSPER\textsubscript{all-norm}  & 17.1 & 39.7 & 67.2 & 87.2 \\
PROSPER\textsubscript{d-norm}   & 16.1 & 42.3 & 71.7 & 91.1 \\
\bottomrule
\end{tabular}
\end{table}

\heading{Offline efficiency analysis}
We compare index size and matching time:
\begin{enumerate*}[label=(\roman*)]
\item On the 1-million-item Multi-CPR corpus, PROSPER uses a 1.5 GB index, compared with about 3 GB for BERT-based dense models and 8 GB for LLM-based dense models;
\item PROSPER retrieves results for 1k queries in under 1 minute, compared with over 3 minutes for Faiss-based dense retrieval.
\end{enumerate*}

\heading{Offline case study}
Table~\ref{tab:hallucination} compares the top-weighted terms learned by Qwen-based SPLADE and PROSPER. For the query \begin{CJK}{UTF8}{gbsn}“爱立舍机油滤芯”\end{CJK} (Elysee oil filter), Qwen-based SPLADE mainly focuses on general oil-filter related terms, while the brand term ``Elysee,'' which reflects the user's specific need, receives insufficient attention. PROSPER instead assigns higher weights to the literal brand tokens while preserving useful expansions related to oil filters. Similarly, for the query \begin{CJK}{UTF8}{gbsn}“老捷豹副水箱”\end{CJK} (Old Jaguar water tank), Qwen-based SPLADE introduces noisy animal-related expansions, whereas PROSPER makes the expansion more concentrated on the user intent. These cases show that PROSPER better balances semantic expansion and literal keywords. 

\section{Online Experiments}
\label{sec:online_experiment}
Next, we describe the deployment of PROSPER in a real-world search engine and present an online evaluation and analysis.

\subsection{System deployment}
\begin{figure}[t]
\centering
\includegraphics[width=\linewidth]{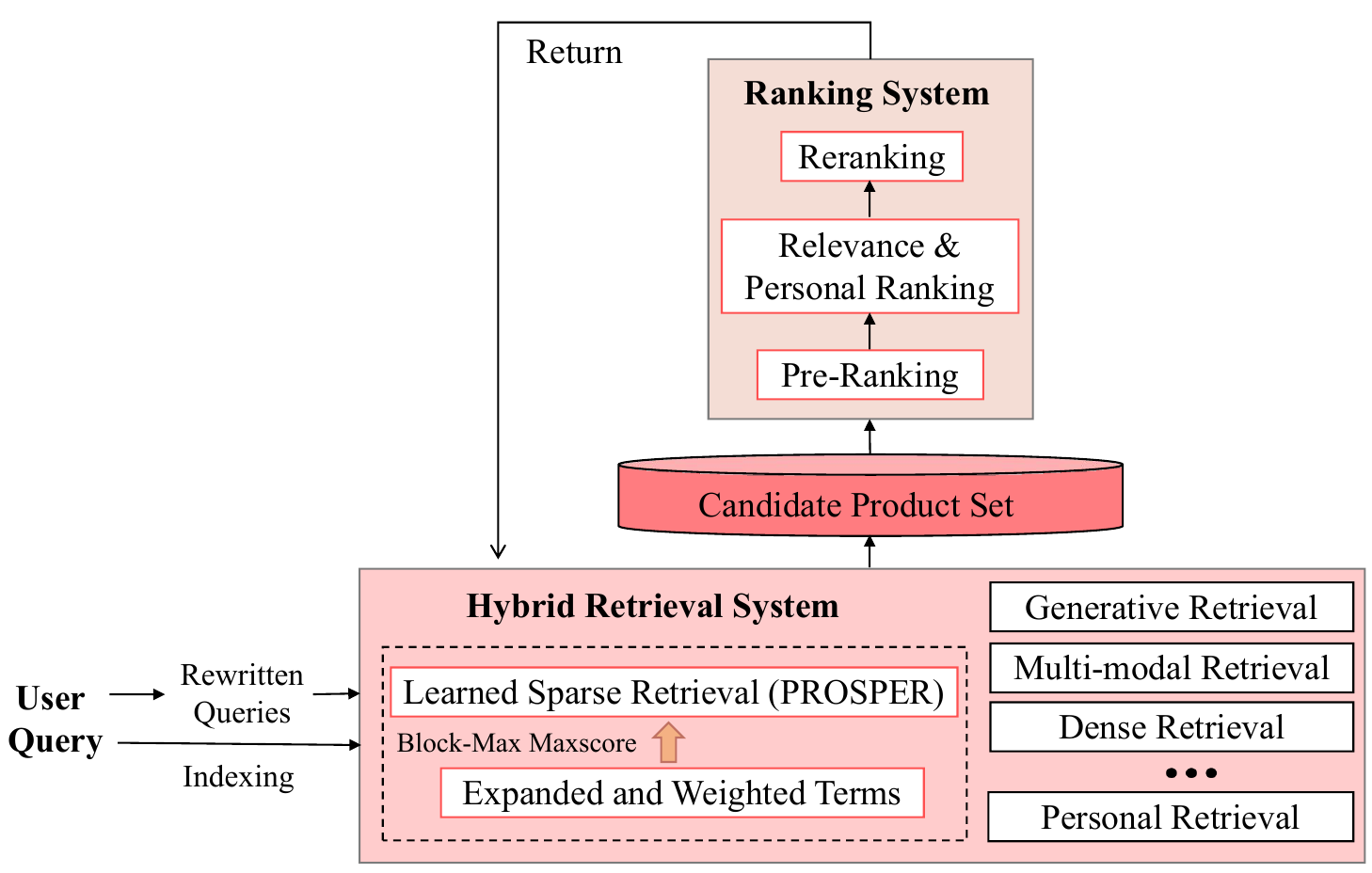}
\caption{Overview of the architecture of the industrial-scale search
engine with the proposed PROSPER.}
\label{fig:system}
\vspace{-2mm}
\end{figure}

Our industrial-scale search engine follows the ``index-retrieve-then-rank'' paradigm \cite{10.1145/3637528.3671654}.
As shown in Figure~\ref{fig:system}, PROSPER is deployed as a sparse retrieval channel in the industrial search engine.
The first-stage product retrieval system combines PROSPER with traditional inverted-index, multi-modal, generative, dense, and personalized retrieval channels to cover diverse user needs.

For first-stage product retrieval, we perform offline inference over the product corpus and associate each product with its weighted terms, which are stored in inverted lists to construct the index. At query time, the query representation provides a set of weighted terms that is used to retrieve relevant products through inverted-index lookup. We use block-max MaxScore \cite{10.1145/2433396.2433412,mallia2024fasterlearnedsparseretrieval,10.1145/3637528.3671654} to improve matching efficiency. The retrieved products then form a candidate set for the subsequent ranking stages.

\subsection{Online experimental setup}
We use Qwen2.5-1.5B as the backbone of the deployed PROSPER model. For training, we sample approximately 330 million user click records collected by the search engine in July 2025 and train the model for one epoch. For deployment, we precompute sparse representations for approximately 80 million high-quality products and 170 million user queries. To improve online matching efficiency, we retain only the 16 highest-weighted query terms for retrieval.

We deploy PROSPER in an online traffic bucket covering approximately 1\% of total traffic and compare it with a baseline bucket using an identical configuration without PROSPER. Other experimental settings follow the offline experiments.

For online evaluation, we divide the business metrics into direct-guided and total-guided metrics. Direct-guided metrics capture purchases made immediately after search, whereas total-guided metrics cover multiple purchase channels, including direct, live-streaming, and advertisement-guided purchases. The evaluated metrics are net gross merchandise volume (GMV), net unique visitors (UV), net transactions, and conversion rate (CVR).

\vspace*{-1mm}
\subsection{Online experimental results}
\textbf{Online A/B test results.}
The one-month A/B test yields improvements in several key business metrics over the baseline bucket, as reported in Table~\ref{tab:online_ab_results}. The CVR also improves by +0.36\%, while other monitored metrics show positive trends. We observe no negative impact on metrics from the ranking, advertising, or recommendation stages. We apply CUPED for variance reduction \cite{10.1145/2433396.2433413}, and the guardrail indicators remain normal.

\begin{table}[t]
\centering
\caption{Online A/B test results.}
\label{tab:online_ab_results}
\begin{tabular}{lcc}
\toprule
\textbf{Metric} & \textbf{Direct guided} & \textbf{Total guided} \\
\midrule
\textbf{Net GMV} & +1.60\% & +1.53\% \\
\textbf{Net Transactions} & +0.31\% & +0.29\% \\
\textbf{Net UV} & +0.24\% & +0.22\% \\
\bottomrule
\end{tabular}
\end{table}

\heading{Online efficiency analysis}
We evaluate online efficiency in terms of index storage and matching latency:
\begin{enumerate*}[label=(\roman*)]
\item For the 80 million products used in our experiment, PROSPER requires a 93 GB index, compared with 152 GB for the dense retrieval channel; 
\item PROSPER has a matching latency of ${\sim}$20 ms, compared with ${\sim}$10 ms for the dense retrieval channel, which benefits from years of dedicated ANN optimization. Both latencies satisfy our production requirements. Moreover, all retrieval channels operate in parallel and therefore do not add serial latency to one another.
\end{enumerate*}

\noindent\textbf{Online case study.}
To further examine the online value of PROSPER, we collect candidate products that are missed by all pre-existing recall channels but retrieved by PROSPER. We evaluate their relevance to the corresponding queries using the online relevance model, followed by manual inspection by domain experts.
\vspace{-1.5mm}
\section{Related Work}
\label{sec:related_work}

\vspace{-.5mm}
\textbf{Dense retrieval.}
Dense retrieval models use neural networks to map text into dense embeddings, achieving strong retrieval performance \cite{guoSemanticModelsFirststage2022,devlin-etal-2019-bert,karpukhin-etal-2020-dense,gaoCondenserPretrainingArchitecture2021}. DPR \cite{karpukhin-etal-2020-dense} introduces a dual-encoder architecture with BERT \cite{devlin-etal-2019-bert}, enabling offline document encoding for efficient retrieval, while ColBERT \cite{10.1145/3397271.3401075} enhances effectiveness through token-level late interaction. Further advances in unsupervised pre-training \cite{izacard2022unsuperviseddenseinformationretrieval,gaoUnsupervisedCorpusAware2022,lei-etal-2023-unsupervised}, hard negative mining \cite{xiong2020approximatenearestneighbornegative}, and knowledge distillation \cite{hofstätter2021improvingefficientneuralranking,10.1145/3404835.3462891} have continued to improve dense retrieval.

Recent work has explored the potential of LLMs for retrieval \cite{wang-etal-2024-improving-text,leeNVEmbedImprovedTechniques2025,taoLLMsAreAlso2024,liuLlama2VecUnsupervisedAdaptation2024,man-etal-2024-ullme}, but their causal attention limits each token to preceding context and thus restricts comprehensive text representation. LLM2Vec \cite{behnamghaderLLM2VecLargeLanguage2024} and NV-Embed \cite{leeNVEmbedImprovedTechniques2025} address this limitation through bidirectional attention and adaptive training, while Echo Embeddings \cite{springerRepetitionImprovesLanguage2024} duplicate inputs to expose tokens to the full context. Nevertheless, dense retrieval offers limited interpretability and incurs high index storage costs.

\heading{Sparse retrieval}
Sparse retrieval \cite{9357643,10.1561/1500000019,10.1145/3404835.3463098} computes relevance using explicit term-level signals. Classical models such as BM25 \cite{10.1561/1500000019} rely on term weighting and inverted indices, while learned sparse retrieval methods use neural networks to assign more informative term weights \cite{10.1145/3269206.3271800,10.1145/3331184.3331303,dai2019contextawaresentencepassagetermimportance,10.1145/3397271.3401204,10.1145/3366423.3380258}. However, their reliance on literal term overlap, combined with the length asymmetry between queries and documents, leads to persistent vocabulary mismatch.
Some methods address this issue by expanding terms next to weighting them \cite{nogueira2019documentexpansionqueryprediction,zhao-etal-2021-sparta,10.1145/3397271.3401262,nogueiraDocumentRankingPretrained2020}; their performance lags behind dense models. SPLADE \cite{10.1145/3404835.3463098}, building on SparTerm \cite{baiSparTermLearningTermbased2020}, optimizes term-weight pooling and introduces FLOPS loss \cite{paria2020minimizingflopslearnefficient} for sparsification. Later versions incorporate hard negatives and distillation, achieving dense-level performance in passage retrieval \cite{formalSPLADEV2Sparse2021,10.1145/3477495.3531857};  follow-up work explores fine-grained query-document interactions \cite{10.1145/3539618.3592065,10.1145/3539618.3591977}.

Top-$k$ masking in SPLADE-mask~\cite{yang2021topk} controls sparsity during training and inference. AdaSparse~\cite{xie2026adasparse} combines learnable soft top-$k$, per-term thresholding, and FLOPS regularization. Our LFW uses a hard top-$k$ window with LRN and FLOPS, which we find improves product retrieval and speeds up sparsification.

Inspired by SPLADE and LLM-based dense retrieval, recent work has begun adapting LLMs for sparse retrieval \cite{doshi2024mistralspladellmsbetterlearned,xuCSPLADELearnedSparse2025,leeNVEmbedImprovedTechniques2025}: \citet{zengScalingSparseDense2025} study the scaling laws of LLM-based sparse retrievers, while CSPLADE \cite{xuCSPLADELearnedSparse2025}, building on NV-Embed \cite{leeNVEmbedImprovedTechniques2025} and Echo Embeddings \cite{springerRepetitionImprovesLanguage2024}, tackles attention-related challenges in applying SPLADE \cite{10.1145/3404835.3463098} to LLMs. However, these studies mainly adapt LLMs through backbone scaling or attention modifications, and their generalization to product search has not been confirmed.
\vspace{-1mm}
\section{Conclusion}
\label{sec:conclusion}

We investigate LLM-based learned sparse retrieval for product search. PROSPER combines a literal residual network (LRN) and a lexical focusing window (LFW) to address lexical expansion hallucination and training initialization, respectively, and achieves improvements in both offline and online evaluation. Our results demonstrate the importance of preserving literal constraints while retaining semantic expansion in product retrieval.

Despite these improvements, using model logits for term expansion and weighting can still introduce noisy terms, and the efficiency of online matching can be further optimized. Future work will explore incorporating CoT reasoning \cite{10.5555/3600270.3602070} and product category information to filter and refine expansion terms. We will also investigate learned sparse retrieval for downstream ranking and other short-text, keyword-sensitive commercial scenarios, including short-video title retrieval and local business search.

\begin{acks}
This work was funded by the Strategic Priority Research Program of the CAS under Grants No. XDB0680102, the National Natural Science Foundation of China (NSFC) under Grants No. 62472408, U25B2076 and 62441229, and the National Key Research and Development Program of China under Grants No. 2023YFA1011602.
This research was also (partially) supported by the Dutch Research Council (NWO), under project numbers 024.004.022, NWA.1389.20.183, and KICH3.LTP.20.006, and the European Union under grant agreement No. 101201510 (UNITE).
All content represents the opinion of the authors, which is not necessarily shared or endorsed by their respective employers and/or sponsors.
\end{acks}

\section*{Ethical Considerations}
The experiments use internal e-commerce click logs and include an online A/B test, so protecting user privacy is especially important. We report only aggregate results and product-search examples and do not release individual user records or personally identifiable information. The internal logs were accessed under the platform's data policies, and the online experiment covered approximately 1\% of traffic with guardrail monitoring to detect unintended effects on the search system.

Generative AI tools were used for language polishing. During case analysis, they served only as a supporting reference while the authors manually inspected whether expanded terms were semantically relevant. They were not used to generate training labels, evaluation ground truth, or reported metrics, and the final analysis was completed and checked by the authors.

\bibliographystyle{ACM-Reference-Format}
\balance
\bibliography{references}

\end{document}